\documentclass[twoside,english,nofootinbib,preprintnumbers,aps,onecolumn,notitlepage]{revtex4}
\usepackage{euscript}
\usepackage{amsthm}
\usepackage{graphicx}
\usepackage{amsfonts}
\usepackage{amsmath}
\usepackage{amssymb}
\usepackage{mathtools}
\usepackage{fancyhdr}
\usepackage{epsfig}
\usepackage{color}
\usepackage{esint}
\usepackage{soul}
\usepackage{calc}
\usepackage{enumerate}
\usepackage{enumitem}
\usepackage{afterpage}
\usepackage{physics}
\usepackage[papersize={8.5in,11in}]{geometry}

\allowdisplaybreaks

\newcommand{\be}{\begin{equation}}
\newcommand{\ee}{\end{equation}}
\begin{document}
\title{Quadrupole formulae with cosmological constant: comparison}

\author{Denis Dobkowski-Ry{\l}ko}
	\email{Denis.Dobkowski-Rylko@ug.edu.pl}
 \affiliation{Institute of Theoretical Physics and Astrophysics, Faculty of Mathematics, Physics and Informatics, University of Gdansk, Wita Stwosza 57, 80-308, Gdańsk, Poland}

\author{Jerzy Lewandowski}
	\email{Jerzy.Lewandowski@fuw.edu.pl}
	\affiliation{Faculty of Physics, University of Warsaw, ul. Pasteura 5, 02-093 Warsaw, Poland}
\begin{abstract}   
We consider three different approaches (by Ashtekar, Bonga and Kesavan; Hoque and Virmani; and Dobkowski-Ryłko and Lewandowski) to investigate gravitational radiation produced by time changing matter source in de Sitter spacetime. All of them lead to generalizations of the quadrupole formula, however, due to different gauge conditions and choices of the hypersurfaces, across which the energy flux is computed, it is nontrivial to see that they all coincide, as one would expect from the symplectic theory. Each of the expressions for the radiated energy in the form of gravitational waves is expressed in terms of the mass and pressure quadrupole moments and written explicitly up to the linear order in $\sqrt{\Lambda}$, or equvalently in Hubble parameter $H$. It is shown that up to the  first order all three of the generalizations of the quadrupole formula agree.

\end{abstract}

\date{\today}


\maketitle

\section{Introduction}
Due to the current observational evidence for very small, however, positive value of the cosmological constant $\Lambda$, there have been many recent developments in the radiation theory for spacetimes with $\Lambda>0$. In particular, there is a set of papers by Ashtekar et al. \cite{ABK1, ABK2, ABK3}, which presents a thorough description of the asymptotics of spacetimes with positive cosmological constant. One of its main results is the generalization of the quadrupole formula for the energy carried by gravitational waves, where the spacelike conformal boundary of de Sitter spacetime was used \cite{ABK3}. 
There have also been other approaches to obtaining the expression for the radiated energy of the compact source, and one of the proposals is to use the cosmological horizon, which is a null surface, as a generalization of the conformal boundary in the radiation theory in Minkowski spacetime. This framework has been investigated, among others, by Houqe and Virmani (HV) \cite{HV} and Dobkowski-Ryłko and Lewandowski (DRL) \cite{DDR&JL}. There are other works, were the asymptotic symmetries and charges in de Sitter spacetime \cite{KL2, ANS, CFR1, PST} and charges and fluxes on cosmological \cite{DGGP}, black hole \cite{DG} and more general, nonexpanding horizons \cite{AKKL1, AKKL2} were studied in the presence of the cosmological constant. The time changing compact source in de Sitter background was also investigated in \cite{Hoque1}, where Isaacson prescription was used in order to define an effective gravitational stress tensor and compute the energy flux. Not only it was shown that the two results of \cite{HV} and \cite{Hoque1} agree, but also that the flux computed on a cosmological horizon agrees with the one evaluated at conformal boundary $\mathcal{I}^+$ of de Sitter spacetime.

The linearized solutions obtained by Ashtekar, Bonga and Kesavan (ABK) \cite{ABK3} in the generalized Lorentz gauge were also derived in Bondi gauge in \cite{CHK}. Moreover, the odd parity quadrupole moments were included in the computations and displacement memory effect have been investigated. The linearized waves obtained in \cite{CHK} were compared to other works on gravitational radiation in de Sitter spacetime \cite{Teukolsky}, in particular by Bonga et al. \cite{BBP} where the Bondi-Sachs framework was used to study the linear as well as non-linear perturbation theory. Despite of different boundary conditions, not only the linearized solutions from \cite{CHK} and \cite{BBP} are shown to be equivalent, but also to the results by Loganayagam and Sheyte \cite{LS} on gravitational perturbations in de Sitter spacetime, where a master variable formalism was applied. Furthermore, a detailed analysis of the asymptotic structure of (anti-)de Sitter  spacetime was performed in partial Bondi gauge in \cite{GZ}. 

Below we briefly present the common part of the derivation of each of the three generalizations of the quadrupole formula considered here. It consists of a set of assumptions used in computations, the general  solution to the linearized Einstein's equations with cosmological constant $\Lambda$ in the presence of a first order linearized source and its form written in terms of the mass and pressure quadrupole moments.

\begin{figure}
    \centering
    \includegraphics[width=0.5\linewidth]{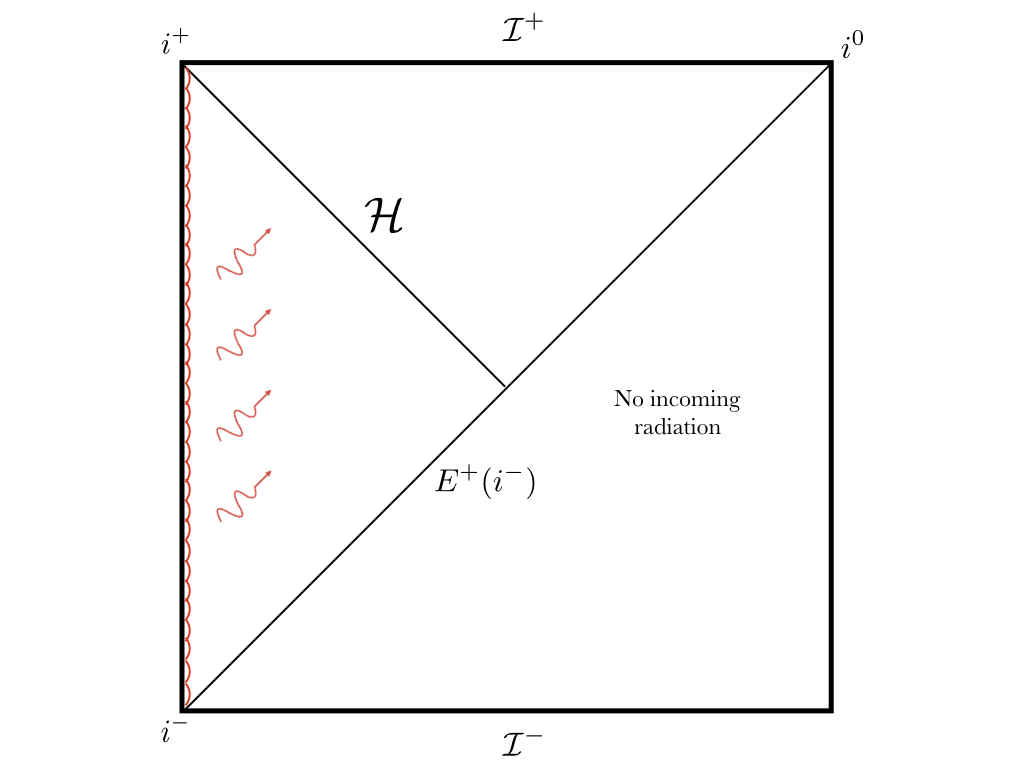}
    \caption{Compact source radiating energy in the form of gravitational waves. Its causal future covers the future Poincar\'e patch (triangle with vertices at $i^-$, $i^0$ and $i^+$). We assume no incoming radiation across the past boundary  $E^+(i^-)$ (future  event horizon of $i^-$) of the future Poincar\'e patch. $\mathcal{H}$ denotes the cosmological horizon.}
    \label{pdiag}
\end{figure}

First, consider a compact, time changing, matter source radiating energy in the form of gravitational waves in de Sitter spacetime. Fig. \ref{pdiag} presents a conformal diagram of the spacetime of interest. Let $\bar g_{\alpha\beta}$ be the background de Sitter metric, defined on the upperleft triangle -- future Poincar\'e patch -- containing the causal future of the source, of the form
\begin{align}\label{etametric}
\bar g_{\alpha\beta} dx^{\alpha}dx^\beta =\tfrac{1}{H^2\eta^2}\mathring g_{\alpha\beta} dx^{\alpha}dx^\beta= \tfrac{1}{H^2\eta^2}(-d\eta^2+dx^2+dy^2+dz^2),
\end{align}
where $H:= \sqrt{\Lambda/3}$ is the Hubble parameter and $\Lambda$ is the cosmological constant. 
Notice, that in this coordinates the cosmological horizon is defined as $\eta+r=0$, whereas on spacelike future infinity $\mathcal{I}^+$ we have $\eta=0$. Consider the first order perturbations of de Sitter spacetime
\begin{align}
g_{\alpha\beta} = \bar g_{\alpha\beta}+\epsilon \gamma_{\alpha\beta},
\end{align}
where $\epsilon$ is the smallness parameter. Similarily as in the radiation theory in Minkowski spacetime, it is convenient to consider the trace-reversed metric perturbation, namely
\begin{align}
\bar \gamma_{\alpha\beta}:=\gamma_{\alpha\beta}-\tfrac{1}{2}\bar g_{\alpha\beta} \gamma.
\end{align}
Consequently, the linearized Einstein's equations in the presence of a first order linearized source $T_{\alpha\beta}$ take the following form
\begin{align}\label{LEE}
 \bar \Box \bar \gamma_{\alpha\beta}-2\bar\nabla_{(\alpha}\bar\nabla^\mu\bar\gamma_{\beta)\mu}+\bar g_{\alpha\beta}\bar\nabla^\mu\bar\nabla^\nu \bar \gamma_{\mu\nu} - \tfrac{2}{3}\Lambda(\bar\gamma_{\alpha\beta}-\bar g_{\alpha\beta}\bar\gamma) = -16\pi T_{\alpha\beta},
\end{align} 
where $\bar \nabla$ and $\bar \Box$ are the derivative and the d'Alembertian operators with respect to the de Sitter metric $\bar g_{\alpha\beta}$.
Equation (\ref{LEE}) have been first derived by Vega et al. \cite{Vega} and also presented more recently in \cite{ABK3}, therefore, here we just provide the final results. Following \cite{ABK3} we introduce a vector field $\eta^\alpha$, which is normal to $\eta=\text{const}$ hypersurface satisfying
\begin{align}
    \eta^\alpha\nabla_\alpha \eta =1,
\end{align}
and choose the gauge condition
\begin{align}\label{gauge1}
\bar\nabla^\alpha \bar\gamma_{\alpha\beta} = 2Hn^\alpha \bar \gamma_{\alpha\beta},
\end{align}
where $n^\alpha:= -H\eta\eta^\alpha$. Then, we decompose the metric perturbation and stress-energy tensor as follows
\begin{align}
\tilde \chi := (\eta^\alpha\eta^\beta+\mathring q^{\alpha\beta})\bar \chi_{\alpha\beta}, && \chi_\alpha :=\eta^\sigma\mathring {q_\alpha}^\beta \bar\chi_{\beta\sigma}, && \chi_{\alpha\beta}:=\mathring {q_\alpha}^\mu \mathring{q_\beta}^\nu \bar \chi_{\mu\nu},\\
 \mathcal{\tilde T}:= (\eta^\alpha\eta^\beta+\mathring q^{\alpha\beta})T_{\alpha\beta}, && \mathcal{T}_\alpha :=\eta^\sigma\mathring {q_\alpha}^\beta  T_{\beta\sigma}, && \mathcal{T}_{\alpha\beta}:=\mathring {q_\alpha}^\mu \mathring{q_\beta}^\nu T_{\mu\nu}.
\end{align}
where $\mathring q^{\alpha\beta}$ is the flat, spacial metric induced by  $\mathring g_{\alpha\beta}$  on a cosmological slice $\eta=\text{const}$, and $\bar\chi_{\alpha\beta}$ is the rescaled trace-reversed metric perturbation defined as $\bar\chi_{\alpha\beta} := H^2\eta^2\bar \gamma_{\alpha\beta}$. Consequently, the solutions to the Einstein's equations may be written in the following form
\begin{align}
\tilde \chi(\eta,\vec x) &= 4\eta \int \frac{d^3\vec x'}{|\vec x-\vec x'|}\frac{1}{\eta_{{}_\text{Ret}}} \mathcal{\tilde T}(\eta_\text{Ret},\vec x') \label{exactsolution1},\\
\chi_{a}(\eta,\vec x) &= 4\eta \int \frac{d^3\vec x'}{|\vec x-\vec x'|}\frac{1}{\eta_{{}_\text{Ret}}}\mathcal{T}_a(\eta_\text{Ret},\vec x') \label{exactsolution2},\\
\chi_{ab}(\eta,\vec x) &= 4\int \frac{d^3 \vec x'}{|\vec x - \vec x'|} \mathcal{T}_{ab}(\eta_{_\text{Ret}},\vec x') + 4\int d^3\vec x' \int_{-\infty}^{\eta_{_\text{Ret}}} d\eta' \frac{1}{\eta'} \partial_{\eta'}\mathcal{T}_{ab}(\eta',\vec x'),\label{exactsolution}
\end{align}
where $\eta_{_\text{Ret}}:=\eta-{|\vec x - \vec x'|}$. Tensors defined on the cosmological slices $\eta=\text{const}$ carry indices denoted by lower Latin letters: $a,b,c, ... = 1,2,3$, as in the expressions above.

Following (\cite{ABK3}) we make several assumption on the radiating isolated system of interest:
\begin{itemize}
\item physical size of the system $D(\eta)$ is uniformly bounded by $D_0$ on all $\eta=\text{const}$ hypersurfaces (which we refer to as cosmological slices);
\item physical size of the system is much smaller than the cosmological radius: $D_0\ll1/H$;
\item system is stationary at distant past and future, namely: $\mathcal{L}_T T_{\alpha\beta}=0$, where $T^\alpha$ is the Killing vector field:
\begin{align}
T^\alpha\partial_\alpha&=-H\Big(\eta\partial_\eta + x\partial_x+ y\partial_y+ z\partial_z\Big);\\
&=\partial_t -Hr\partial_r
\end{align}
\item no incoming radiation across the past boundary  $E^+(i^-)$ (future  event horizon of $i^-$) of the future Poincar\'e patch; 
\item the velocity of the source is small: $v\ll 1$ (we use the units where $c=1=G$).
\end{itemize}
Then, the trace reversed rescaled metric perturbation can be put in the form
\begin{align}\label{solutionRet}
\chi_{ab}(\eta,\vec x) = \frac{4}{r}\int {d^3 \vec x'} \mathcal{T}_{ab}(\eta_{_\text{ret}},\vec x') + 4\int d^3\vec x' \int_{-\infty}^{\eta_{_\text{ret}}} d\eta' \frac{1}{\eta'} \partial_{\eta'}\mathcal{T}_{ab}(\eta',\vec x'),
\end{align}
where $\eta_{\text{ret}}=\eta-r$.

Mass and pressure quadrupole moments are defined as
\begin{align}
Q_{ab}^{(\rho)}(\eta)&:= \int_\Sigma d^3V \rho(\eta) \bar x_a \bar x_b,\label{massquadrupolemoment}\\
Q_{ab}^{(p)}(\eta)&:= \int_\Sigma d^3V \big(p_1(\eta)+p_2(\eta)+p_3(\eta)\big) \bar x_a \bar x_b,\label{pressurequadrupolemoment}
\end{align}
where $\Sigma$ is any cosmological slice, $d^3V$ denotes proper volume element and $\bar x_a:= -\tfrac{1}{H\eta}x_a$. Matter density $\rho$ and pressure $p_i$ are defined as usual, that is
\begin{align}
\rho = T_{\alpha\beta}n^\alpha n^\beta,&& p_i=T^{ab}\partial_a x_i \partial_b x_i.
\end{align}
Retarded solution (\ref{solutionRet}) may be expressed in terms of the quadrupole moments in a similar way as in Minkowski case (for details see \cite{ABK3})
\begin{align}
\chi_{ab}&=-\tfrac{2H}{r}\eta_\text{ret}[\ddot Q_{ab}^{(\rho)}+2H\dot Q_{ab}^{(\rho)}+H\dot Q_{ab}^{(p)}+2H^2Q_{ab}^{(p)}]({\eta_\text{ret}})\nonumber\\
&-2H\int^{\eta_\text{ret}}_{-\infty} \frac{d\eta'}{\eta'}\partial_\eta' \bigg(\eta'[\ddot Q_{ab}^{(\rho)}+2H\dot Q_{ab}^{(\rho)}+H\dot Q_{ab}^{(p)}+2H^2 Q_{ab}^{(p)}](\eta')\bigg)
\end{align}
where the dot corresponds to the lie derivative with respect to the time translation Killing vector field $T$.

\section{Comparison of the generalizations of quadrupole formulae}
Over the last decade, the generalization of the radiation framework for spacetimes with positive cosmological constant attracted attention of many relativists. In \cite{ABK3} found a generalized quadrupole formula on future infinity $\mathcal{I}^+$ in the TT-gauge was found (ABK). In (HV) and (DRL) approach, on the other hand, the cosmological Killing horizon, which is a null surface,  was used as the conformal boundary from Minkowski spacetime. Then, the energy flux across such generalization of the conformal boundary was calculated. However, despite the fact, that the same null surface has been chosen, the two derivations differ and the resulting expressions are not manifestly equivalent. In particular, (HV) uses the symplectic current density of the covarian phase space approach, that was introduced in \cite{WH}, together with the tt-projection in their derivation of the quadrupole formula, whereas (DDR) concentrates on Wald-Zoupas \cite{WZ} and Chandrasekaran-Flanagan-Prabhu \cite{CFP} theory of radiation through the null surfaces has been applied and a generalization of the quadrupole formula has been derived up to linear terms in the Hubble parameter $H$. This allowed to explicitly show that in the limit when $\Lambda\rightarrow 0$ the result recovers Einstein's quadrupole formula. The goal of the current paper is to check whether the three expressions for generalized quadrupole formulae agree, at least, up to linear order terms in $H$ (or equivalently $\sqrt{\Lambda}$).

\subsection{Generalization of the quadrupole formula by Dobkowski-Ry\l{}ko and Lewandowski}

We begin with (DRL) as in this approach the result is already written in the desired form, namely, such that allows to take a limit as $\Lambda$ approaches zero just by setting $\sqrt{\Lambda}=0$ (or $H=0$). Since $(\eta, \vec x)$ coordinates are not suitable to take such limit, one needs to move to a different coordinate system. Therefore, we introduce coordinates $(t, r, \theta, \varphi)$, which are defined as
\begin{align}
\eta= -\tfrac{1}{H}e^{-Ht}, && x = r \cos\theta, && y = r \sin\theta\cos\varphi, &&z = r\sin\theta\sin\varphi,
\end{align}
in which the metric tensor $\bar g_{\alpha\beta}$ takes the following form
\begin{align}
\bar g_{\alpha\beta} dx^{\alpha}dx^\beta = -dt^2 +e^{2Ht}(dr^2+r^2d\theta^2+r^2\sin^2\theta d\varphi^2).
\end{align}
Notice, that now, by setting $H=0$, the metric coefficients simplify to those of the Minkowski metric. In (DRL) approach, authors used the general formula for the energy flux through a null surface derived by Chandrasekaran et al. \cite{CFP}. It reads
\begin{align}\label{energy1}
E_T &= \frac{1}{8\pi}\int_{ \mathcal{H}} d^3V \bigg(\sigma_{AB}\sigma^{AB}-\tfrac{1}{2}\theta^2 \bigg)
\end{align}
where $\sigma_{AB}$ and $\theta$ are shear and expansion, respectively, of the Killing vector field $T^\alpha$, whereas $\mathcal{H}$ is the cosmological horizon with the proper volume form $d^3V$. A suitable gauge condition needs to be implemented, for the formula (\ref{energy1}) to be applicable. 
Namely, the perturbed cosmological horizon defined by $r=\exp(-Ht)/H$ must remain null\footnote{For details see \cite{DDR&JL}.}. Consequently, a generalization of the quadrupole formula has been derived up to terms linear in $H$
\begin{align}\label{quadrupoleDRL}
E_T=\frac{1}{45}\int dt\sum_{i,j=1}^3\Bigg[\bigg( \frac{d^3q^{(\rho)}_{ij}}{dt^3}\bigg)^2+2H\Bigg( \frac{d^3q^{(\rho)}_{ij}}{dt^3}\bigg(\frac{d^2q^{(p)}_{ij}}{dt^2}{-7}\frac{d^2q^{(\rho)}_{ij}}{dt^2}\bigg)\Bigg)\Bigg](t_{\text{ret}})+\mathcal{O}(H^2), 
\end{align}
where
\begin{align}\label{tracelessQM}
q^{(i)}_{ab}:=&\mathring q^{ab}Q^{(i)}_{ab}-\tfrac{1}{3}\mathring q_{ab}Q^{(i)},\\
\end{align}
are the traceless quadrupole moments and $i$ stands for $\rho$ (mass quadrupole moment) or $p$ (pressure quadrupole moment). Here the retared time $t_\text{ret}$ is defined as:
\begin{align}
    t_\text{ret} =& -\frac{1}{H}\ln (e^{-Ht}+Hr) =  -\frac{1}{H}\ln (2e^{-Ht})
\end{align}
and the last equality holds on the horizon $\mathcal{H}$. Expression (\ref{quadrupoleDRL}) is invariant under transformations $t\rightarrow t+\frac{1}{H}c$, where $c=\text{const}$. By setting $c=\ln 2$, we obtain the same expression as (\ref{quadrupoleDRL}), but now with the square bracket evalutated at $t$ instead of $t_\text{ret}$. Furthermore, notice, that the last term in the square bracket in (\ref{quadrupoleDRL}) maybe in turned into the total derivative, namely
\begin{align}\label{totalD}
\frac{d^3q^{(\rho)}_{ij}}{dt^3}\frac{d^2q^{(\rho)}_{ij}}{dt^2} = \frac{1}{2}\frac{d}{dt}\bigg( \frac{d^2q^{(\rho)}_{ij}}{dt^2}\bigg)^2,
\end{align}
which integrated vanishes for the  source of our interest -- stationary at distant past and future. Therefore, this version of the generalized quadrupole formula simplifies to
\begin{align}\label{quadrupoleDDR&JLsimlified}
    E_T=\frac{1}{45}\int dt\sum_{i,j=1}^3\Bigg[\bigg( \frac{d^3q^{(\rho)}_{ij}}{dt^3}\bigg)^2+2H\bigg( \frac{d^3q^{(\rho)}_{ij}}{dt^3}\frac{d^2q^{(p)}_{ij}}{dt^2}\bigg)\Bigg]+\mathcal{O}(H^2).
\end{align}
Taking a limit for the vanishing cosmological constant is straightforward and requires only setting $H$ to zero, the result reads
\begin{align}
    E_T=\frac{1}{45}\int dt\sum_{i,j=1}^3\bigg( \frac{d^3q^{(\rho)}_{ij}}{dt^3}\bigg)^2
\end{align}
and coincides with the famous Einstein quadrupole formula \cite{Wald}.

\vspace{5mm}

\subsection{Generalization of the quadrupole formula by Hoque and Virmani}
In this subsection we investigate the (HV) version of the generalized quadrupole formula, which is also an expression for energy flux across the cosmological horizon. It was shown that, by fully exhausting gauge freedom, one may obtain the following constraint
\begin{align}
    \partial^i\chi_{ij} = {\chi^i}_i = 0 = \chi_{00}.
\end{align}
To extract the transverse traceless part of the metric perturbation, the tt-projection is applied:
\begin{align}
\chi^{\text{tt}}_{ij} = {\Lambda_{ij}}^{kl}\chi_{kl}
\end{align}
where:
\begin{align}
{\Lambda^{kl}}_{ij} &= \tfrac{1}{2}({P_i}^k{P_j}^l+{P_i}^l{P_j}^k-P_{ij}P^{kl}),\\
{P_i}^j&={\delta_i}^j-\tilde x_i \tilde x^j,\\\label{xTilde}
\tilde x^i &= x^i/r.
\end{align}
In \cite{HV} it is stressed that the radial tt-projection is not an appropriate operation to extract transverse-traceless part of the metric perturbation $\chi_{ij}$, however, it is expected that for the radial distances far from the source, it seems reasonable that the tt-projection will provide good results. The correct notion to analyse asymptotically de Sitter spacetimes is the TT-gauge which is provided by the decomposition of $\chi_{ij}$, namely
\begin{align}
    \chi_{ij} = \frac{1}{3}\delta_{ij}\delta^{kl}\chi_{kl}+\Big(\partial_i \partial_j - \tfrac{1}{3}\delta_{ij}\nabla^2 \Big)B+\partial_iB_j^T+\partial_jB^T_i+\chi_{ij}^{TT}.
\end{align}
In the above decomposition $B_i^T$ is transverse, that is $\partial^iB_i^T=0$, and $\chi_{ij}^{TT}$ is the gauge invariant term, which is transverse-traceless
\begin{align}
\partial^i\chi_{ij}^{TT}=\delta^{ij}\chi_{ij}^{TT} = 0.
\end{align}
For more details on the conceptual difference of the tt-projection and TT-gauge see \cite{AB}. The final expression for the energy flux across the cosmological horizons $\mathcal{H}$ obtained by (HV) approach is of the form
\begin{align}\label{quadrupoleHV}
E_T = \frac{1}{8\pi} \int d\tau \int_{S_2}d\Omega \ R^{\text{tt}}_{ij} \ R^{\text{tt}}_{kl}\ \delta^{ik}\delta^{jl},
\end{align}
where:
\begin{align}\label{Rij0}
R_{ij} &= \dot A_{ij}+HA_{ij} = T^\mu\partial_\mu A_{ij} - HA_{ij},\\
A_{ij}&= \ddot Q^{(\rho)}_{ij} +2H\dot Q^{(\rho)}_{ij}+H\dot Q^{(p)}_{ij}+2H^2 Q^{(p)}_{ij},
\end{align}
and the dots indicate the Lie derivatives with respect to the Killing vector field $T^\mu$, namely $\dot X_{ab}:=\mathcal{L}_TX_{ab}$, whereas $\tau$ is the Killing parameter along the integral curves of $T^\mu$. We choose $(\tau, \theta, \varphi)$ to be the coordinates on the the cosmological horizon $\mathcal{H}$, such that
\begin{align}
    \frac{\partial\eta}{\partial\tau} = - H\eta, && \frac{\partial x^i}{\partial\tau}=-Hx^i.
\end{align}
Consequently, it is convenient  to express $T^\mu\partial_\mu$ as a derivative with respect to $\tau$
\begin{align}
    A_{ij}&= \frac{d^2}{d\tau^2} Q^{(\rho)}_{ij} -2H\frac{d}{d\tau}Q^{(\rho)}_{ij}+H\frac{d}{d\tau} Q^{(p)}_{ij},\\\label{Rij}
    R_{ij} & = \frac{d^3}{d\tau^3} Q^{(\rho)}_{ij} -2H\frac{d^2}{d\tau^2}Q^{(\rho)}_{ij}+H\frac{d^2}{d\tau^2} Q^{(p)}_{ij} -H\frac{d^2}{d\tau^2} Q^{(\rho)}_{ij} +2H^2\frac{d}{d\tau}Q^{(\rho)}_{ij}-H^2\frac{d}{d\tau} Q^{(p)}_{ij}\nonumber\\
    & = \frac{d^3}{d\tau^3} Q^{(\rho)}_{ij} -3H\frac{d^2}{d\tau^2}Q^{(\rho)}_{ij}+H\frac{d^2}{d\tau^2} Q^{(p)}_{ij}  +\mathcal{O}(H^2).
\end{align}

We start by computing the expression for $R^{\text{tt}}_{ij}$ in terms of $R_{ij}$ and $R=\delta^{ij}R_{ij}$ that is
\begin{align}
R^{\text{tt}}_{ij} &= {\Lambda_{ij}}^{kl}R_{kl}\\
&=R_{ij}-R_{il}\tilde x_j \tilde x^l-R_{kj}\tilde x_i \tilde x^k+\tfrac{1}{2}R_{kl}\tilde x_i \tilde x^k\tilde x_j \tilde x^l-\tfrac{1}{2}\delta_{ij}(R-R_{kl}\tilde x^k \tilde x^l)+\tfrac{1}{2}R\tilde x_i \hat x_j.
\end{align}
We may then expand the expression under the integral in (\ref{quadrupoleHV}) to
\begin{align}
R^{\text{tt}}_{ij} \ R^{\text{tt}}_{kl} \delta^{ik}\delta^{jl} &= R^{\text{tt}}_{xx} \ R^{\text{tt}}_{xx} +2R^{\text{tt}}_{xy} \ R^{\text{tt}}_{xy} +2R^{\text{tt}}_{xz} \ R^{\text{tt}}_{xz} +R^{\text{tt}}_{yy} \ R^{\text{tt}}_{yy}+2R^{\text{tt}}_{yz} \ R^{\text{tt}}_{yz}+R^{\text{tt}}_{zz} \ R^{\text{tt}}_{zz}.
\end{align}
Each term on the right hand side has to be integrated over the cosmological horizon $\mathcal{H}$. Here we show how to perform the integration of $R^\text{tt}_{xx}R^\text{tt}_{xx}$, namely
\begin{align}
\frac{1}{8\pi} \int d\tau \int_{S_2}d\Omega R^{\text{tt}}_{xx} \ R^{\text{tt}}_{xx}&=\frac{1}{24} \int d\tau (\tfrac{128}{105}R^2_{xx}+\tfrac{4}{7}R^2_{yy}+\tfrac{4}{7}R^2_{zz}+\tfrac{32}{105}R^2_{xy}+\tfrac{32}{105}R^2_{xz}+\tfrac{16}{15}R^2_{yz}\nonumber\\
&-\tfrac{128}{105}R_{xx}R_{yy}-\tfrac{128}{105}R_{xx}R_{zz}+\tfrac{8}{105}R_{yy}R_{zz}).
\end{align}
In a similar fashion we find the other terms and sum them up. The result reads
\begin{align}
\frac{1}{8\pi} \int d\tau \int_{S_2}d\Omega \ R^{\text{tt}}_{ij} \ R^{\text{tt}}_{kl} \delta^{ik}\delta^{jl} &=\tfrac{2}{15}\int d\tau \Big(R^2_{xx}+R^2_{yy}+R^2_{zz}+3(R^2_{xy}+R^2_{xz}+R^2_{yz})\nonumber\\
&-R_{xx}R_{yy}-R_{xx}R_{zz}-R_{yy}R_{zz}\Big).
\end{align}
Using the expression (\ref{Rij}) to express $R_{ij}$ in terms of the mass and pressure quadrupole moments yields:
\begin{align}\label{energyfinal0}
E_T&=\tfrac{2}{15}\int d\tau   \Bigg[(\partial_\tau^3 Q_{xx}^{(\rho)})^2+(\partial_\tau^3 Q_{yy}^{(\rho)})^2+(\partial_\tau^3 Q_{zz}^{(\rho)})^2+3(\partial_t^3 Q_{xy}^{(\rho)})^2+3(\partial_t^3 Q_{xz}^{(\rho)})^2+3(\partial_t^3 Q_{yz}^{(\rho)})^2\nonumber\\
&-(\partial_\tau^3 Q_{xx}^{(\rho)})(\partial_\tau^3 Q_{yy}^{(\rho)})-(\partial_\tau^3 Q_{xx}^{(\rho)})(\partial_t^3 Q_{zz}^{(\rho)})-(\partial_\tau^3 Q_{yy}^{(\rho)})(\partial_\tau^3 Q_{zz}^{(\rho)})\nonumber\\
&+2H\partial_\tau^3 Q_{xx}^{(\rho)}\Big(\partial_\tau^2 Q_{xx}^{(p)}-3\partial_\tau^2 Q_{xx}^{(\rho)}\Big)+2H\partial_\tau^3 Q_{yy}^{(\rho)}\Big(\partial_\tau^2Q_{yy}^{(p)}-3\partial_\tau^2 Q_{yy}^{(\rho)}\Big)\nonumber\\
&+2H\partial_\tau^3 Q_{zz}^{(\rho)}\Big(\partial_\tau^2Q_{zz}^{(p)}-3\partial_\tau^2 Q_{zz}^{(\rho)}\Big)+6H\partial_\tau^3 Q_{xy}^{(\rho)}\Big(\partial_\tau^2Q_{xy}^{(p)}-3\partial_\tau^2 Q_{xy}^{(\rho)}\Big)\nonumber\\
&+6H\partial_\tau^3 Q_{xz}^{(\rho)}\Big(\partial_\tau^2Q_{xz}^{(p)}-3\partial_\tau^2 Q_{xz}^{(\rho)}\Big)+6H\partial_\tau^3 Q_{yz}^{(\rho)}\Big(\partial_\tau^2Q_{yz}^{(p)}-3\partial_\tau^2 Q_{yz}^{(\rho)}\Big)\nonumber\\
&-H\partial_\tau^3 Q_{xx}^{(\rho)}\Big(\partial_\tau^2Q_{yy}^{(p)}-3\partial_\tau^2 Q_{yy}^{(\rho)}\Big)-H\partial_\tau^3 Q_{yy}^{(\rho)}\Big(\partial_\tau^2Q_{xx}^{(p)}-3\partial_\tau^2 Q_{xx}^{(\rho)}\Big)\nonumber\\
&-H\partial_\tau^3 Q_{xx}^{(\rho)}\Big(\partial_\tau^2Q_{zz}^{(p)}-3\partial_\tau^2 Q_{zz}^{(\rho)}\Big)-H\partial_\tau^3 Q_{zz}^{(\rho)}\Big(\partial_\tau^2Q_{xx}^{(p)}-3\partial_\tau^2 Q_{xx}^{(\rho)}\Big)\nonumber\\
&-H\partial_\tau^3 Q_{yy}^{(\rho)}\Big(\partial_\tau^2Q_{zz}^{(p)}-3\partial_\tau^2 Q_{zz}^{(\rho)}\Big)-H\partial_\tau^3 Q_{zz}^{(\rho)}\Big(\partial_\tau^2Q_{yy}^{(p)}-3\partial_\tau^2 Q_{yy}^{(\rho)}\Big)\Bigg]+\mathcal{O}(H^2).
\end{align}
Furthermore, similarly as in \cite{DDR&JL}, we introduce the traceless quadrupole moments $q^{(i)}_{ab}$ via (\ref{tracelessQM}) to simplify the above expression. 

Finally, the generalized quadrupole formula by Hoque et al. (\ref{quadrupoleHV}) expanded in the Hubble parameter and written up to linear order in $H$ is of the form
\begin{align}\label{HVquadFinal}
E_T=\tfrac{1}{45}\int d\tau\sum_{i,j=1}^3\Bigg[\bigg( \tfrac{d^3q^{(\rho)}_{ij}}{d\tau^3}\bigg)^2+2H\bigg( \tfrac{d^3q^{(\rho)}_{ij}}{d\tau^3}\Big(\tfrac{d^2q^{(p)}_{ij}}{d\tau^2}{-3}\tfrac{d^2q^{(\rho)}_{ij}}{d\tau^2}\Big)\bigg)\Bigg]+\mathcal{O}(H^2).
\end{align}
Notice, that here we may also express the last term in the square bracket above as a total derivative (as in (\ref{totalD}) but with a different constant factor), which after integration yields the boundary terms that vanish due to the assumption of the stationarity of the source in far past and future. Consequently, the result coincides with the formula (\ref{quadrupoleDDR&JLsimlified}) found in \cite{DDR&JL} up to the linear order terms in Hubble parameter $H = \sqrt{\Lambda/3}$.

\subsection{Generalization of the  quadrupole formula by Ashtekar, Bonga and Kesavan}
The last energy flux formula for gravitational radiation in de Sitter background that we consider in this paper was
calculated in (ABK) approach at future infinity $\mathcal{I}^+$ and is of the form
\begin{align}\label{AshtekarQF}
E_T \hat = \frac{1}{16\pi H} \int_{\mathcal{I}^+}d^3x \mathcal{E}_{cd}\big( \mathcal{L}_T\chi_{ab}+2H\chi_{ab} \big)\mathring q^{ac}\mathring q^{bd},
\end{align}
where "$\hat =$" denotes the value at $\mathcal{I}^+$ and
\begin{align}\label{electricWeyl}
\mathcal{E}_{ab} = \frac{1}{2H\eta}\bigg( \mathring q_a^c \mathring q_b^d - \tfrac{1}{3}\mathring q_{ab} \mathring q^{cd}\bigg) \bigg[\tfrac{1}{2} \mathring D_c\mathring D_d \tilde \chi-\mathring D_{(c}\mathring D^m\chi_{d)m} -\mathring D_{(c}\partial_\eta\chi_{d)}+\Big(\partial_\eta^2-\tfrac{1}{\eta}\partial_\eta \Big)\chi_{cd} \bigg]
\end{align}
is the perturbed electric part of the Weyl curvature. It can be shown, that due to the gauge condition (\ref{gauge1}) the above expression for $\mathcal{E}_{ab}$ simplifies to
\begin{align}
    \mathcal{E}_{ab} = \frac{1}{2H\eta}\bigg( \partial_\eta^2-\frac{1}{\eta}\bigg)\chi_{ab}^{TT}.
\end{align}
Furthermore, the energy flux across future infinity $\mathcal{I}^+$ simplifies to:
\begin{align}\label{quadrupoleTTABK}
    E_T\hat = \frac{1}{8\pi}\int_\mathcal{I^+}dTd^2S \Big[ R_{ab}R_{cd}^{TT}\mathring q^{ac}\mathring q^{bd}\Big],
\end{align}
where $T$ is the affine parameter along integral curves of the Killing vector field $T^a$ satisfying $dT=-dr/(rH)$ at future infinity and the radiation field $R_{ab}$ is again of the form (\ref{Rij0}). Due to the fact that extracting TT part of a tensor is a highly nontrivial procedure, instead of expanding eq. (\ref{quadrupoleTTABK}) in Hubble parameter, we will use the intermediate form of the quadrupole formula (\ref{semiAshtekarQF}). 

Therefore, we  proceed by calculating each term present in the expression (\ref{electricWeyl}) for the electric part of Weyl tensor $\mathcal{E}_{ab}$. Gauge condition (\ref{gauge1}) can be split into two separate equations of the following form
\begin{align}
\mathring D^a \chi_{ab} = \partial_\eta \chi_b - \tfrac{2}{\eta} \chi_b, && \mathring D^a \chi_a = \partial_\eta\big( \tilde \chi - \chi \big) -\tfrac{1}{\eta}\tilde \chi,\nonumber
\end{align}
where $\chi=\mathring q^{ab}\chi_{ab}$. These equations may be solved and provide the expressions for $\chi_a$ and $\tilde \chi$ written in terms of the space-space component of the metric perturbation $\chi_{ab}$, namely
\begin{align}\label{tildeChi}
\tilde \chi &= \eta \int^\eta_{-\infty} \frac{\mathring D^a\chi_a+\partial_{\eta'} \chi}{\eta'} d\eta',\\ \label{ChiA}
\chi_a &= \eta^2 \int^\eta_{-\infty} \frac{\mathring D^a\chi_{ab}}{\eta'^2}d\eta'.
\end{align}
For the convenience we introduce the following notation
\begin{align}
\mathbb{T}_{ab}(\eta):= \int {d^3 \vec x'} \mathcal{T}_{ab}(\eta,\vec x'),
\end{align}
which allows to write the metric  perturbation $\chi_{ab}$ in a more compact form
\begin{align}\label{solution}
\chi_{ab}(\eta,\vec x) = \frac{4}{r}\mathbb{T}_{ab}(\eta_\text{ret})+  4 \int_{-\infty}^{\eta_{_\text{ret}}} d\eta' \frac{1}{\eta'} \partial_{\eta'}\mathbb{T}_{ab}(\eta').
\end{align}

We begin with calculating the divergence of the metric perturbation $\chi_{ab}$ with respect to the covariant derivative $\mathring D_a$ (of the spatial metric $\mathring q_{ab}$)
\begin{align}\label{Dchi}
\mathring D^a\chi_{ab}&=  -\frac{4x^a}{r^3}\mathbb{T}_{ab}(\eta_\text{ret})+ \frac{4}{r}D^a\mathbb{T}_{ab}(\eta_\text{ret})+ 4D^a \int_{-\infty}^{\eta_{_\text{ret}}} d\eta' \frac{1}{\eta'} \partial_{\eta'}\mathbb{T}_{ab}(\eta')\nonumber\\
&=  -\frac{4x^a}{r^3}\mathbb{T}_{ab}(\eta_\text{ret})- \frac{4x^a\eta}{r^2\eta_\text{ret}}\partial_\eta\mathbb{T}_{ab}(\eta_\text{ret}).
\end{align}
Then, we input the above to the eq. (\ref{ChiA}) and find
\begin{align}
\chi_b &= -\eta^2 \int^\eta_{-\infty} \bigg(\frac{4x^a}{\eta'^2r^3}\mathbb{T}_{ab}(\eta'_\text{ret})+ \frac{4x^a}{r^2\eta\eta_\text{ret}}\partial_\eta\mathbb{T}_{ab}(\eta'_\text{ret})\bigg)d\eta'\nonumber\\
&= -\frac{4x^a\eta^2}{r^2} \int^\eta_{-\infty} \bigg(\frac{1}{\eta'^2r}\mathbb{T}_{ab}(\eta'_\text{ret})-\frac{r-2\eta'}{\eta'^2\eta'^2_\text{ret}}\mathbb{T}_{ab}(\eta'_\text{ret})\bigg)d\eta' -\frac{4x^a\eta}{r^2{\eta_\text{ret}}} \mathbb{T}_{ab}(\eta_\text{ret})\nonumber\\
&= -\frac{4x^a\eta^2}{r^3} \int^\eta_{-\infty} \frac{1}{\eta'^2_\text{ret}}\mathbb{T}_{ab}(\eta'_\text{ret})d\eta' -\frac{4x^a\eta}{r^2{\eta_\text{ret}}} \mathbb{T}_{ab}(\eta_\text{ret}),
\end{align}
where in the second line we integrated by parts. Next, we calculate the partial derivative with respect to $\eta$ of the above expression, which yields:
\begin{align}\label{partial_etaChiA}
\partial_\eta\chi_b &=-\frac{8x^a\eta}{r^3} \int^\eta_{-\infty} \frac{1}{\eta'^2_\text{ret}}\mathbb{T}_{ab}(\eta'_\text{ret})d\eta'-\frac{4(\eta+r)x^a}{r^3\eta_\text{ret}} \mathbb{T}_{ab}(\eta_\text{ret}) -\frac{4x^b\eta}{\eta_\text{ret}r^2}\partial_\eta\mathbb{T}_{ab}(\eta_\text{ret}).
\end{align}
Then, by adding equations (\ref{Dchi}) and (\ref{partial_etaChiA}) we obtain
\begin{align}
\mathring D^a\chi_{ab}+\partial_\eta\chi_b&=  -\frac{8x^a\eta}{r^3\eta_\text{ret}}\mathbb{T}_{ab}(\eta_\text{ret})- \frac{8x^a\eta}{r^2\eta_\text{ret}}\partial_\eta\mathbb{T}_{ab}(\eta_\text{ret})-\frac{8x^a\eta}{r^3} \int^\eta_{-\infty} \frac{1}{\eta'^2_\text{ret}}\mathbb{T}_{ab}(\eta'_\text{ret})d\eta'.
\end{align}
Finally, we calculate the covariant derivative of the above expression to find the middle two terms in the square bracket of eq. (\ref{electricWeyl}), which yields
\begin{align}\label{middleElectricWeyl}
\mathring D_c(D^a\chi_{ab}+\partial_\eta\chi_b)&=  -\frac{8\eta}{r^3\eta_\text{ret}}\mathbb{T}_{bc}(\eta_\text{ret})+\frac{24x^ax_c\eta}{r^5\eta_\text{ret}}\mathbb{T}_{ab}(\eta_\text{ret})+\frac{8(3\eta-4r)\eta x^ax_c}{r^4\eta^2_\text{ret}}\partial_\eta\mathbb{T}_{ab}(\eta_\text{ret})\nonumber\\
&- \frac{8\eta}{r^2\eta_\text{ret}}\partial_\eta\mathbb{T}_{bc}(\eta_\text{ret})+ \frac{8x_cx^a\eta}{r^3\eta_\text{ret}}\partial^2_\eta\mathbb{T}_{ab}(\eta_\text{ret})-\frac{8\eta}{r^3} \int^\eta_{-\infty} \frac{1}{\eta'^2_\text{ret}}\mathbb{T}_{bc}(\eta'_\text{ret})d\eta'\nonumber\\
&+\frac{24x_cx^a\eta}{r^5} \int^\eta_{-\infty} \frac{1}{\eta'^2_\text{ret}}\mathbb{T}_{ab}(\eta'_\text{ret})d\eta'.
\end{align}
Notice the overall factor of $1/\eta$ in (\ref{electricWeyl}), therefore, before evaluating the value of (\ref{middleElectricWeyl}) at $\mathcal{I}^+$, we have to first divide it by $\eta$ and then set $\eta=0$, namely
\begin{align}\label{1/etamiddleElectricWeyl}
\frac{1}{\eta}\mathring D_c(D^a\chi_{ab}+\partial_\eta\chi_b)&\hat =  \frac{8}{r^4}\mathbb{T}_{bc}(\eta_\text{ret})-\frac{24x^ax_c}{r^6}\mathbb{T}_{ab}(\eta_\text{ret})-\frac{32 x^ax_c}{r^5}\partial_\eta\mathbb{T}_{ab}(\eta_\text{ret})\nonumber\\
&+ \frac{8}{r^3}\partial_\eta\mathbb{T}_{bc}(\eta_\text{ret})- \frac{8x_cx^a}{r^4}\partial^2_\eta\mathbb{T}_{ab}(\eta_\text{ret})-\frac{8}{r^3} \int^0_{-\infty} \frac{1}{\eta'^2_\text{ret}}\mathbb{T}_{bc}(\eta'_\text{ret})d\eta'\nonumber\\
&+\frac{24x_cx^a}{r^5} \int^0_{-\infty} \frac{1}{\eta'^2_\text{ret}}\mathbb{T}_{ab}(\eta'_\text{ret})d\eta'.
\end{align}
Moreover, since we are interested in expanding (\ref{AshtekarQF}) to zeroth and first order terms in Hubble parameter $H$, it is sufficient only to consider the leading order terms in $1/r$ of (\ref{1/etamiddleElectricWeyl}), which will become clear when we use the affine parameter of the integral curves of $T^a$ as a coordinate on $\mathcal{I}^+$. With that in mind, we compute
\begin{align}\label{leadingMiddleElectricWeyl}
\frac{1}{\eta}\mathring D_c(D^a\chi_{ab}+\partial_\eta\chi_b)&\hat=- \frac{8x_cx^a}{r^4}\partial^2_\eta\mathbb{T}_{ab}(\eta_\text{ret})+\mathcal{O}(H^3).
\end{align}
Here, we will not calculate the remaining two terms of (\ref{electricWeyl}) explicitly, but only present the leading order terms in Hubble parameter $H$ of their final expressions (for a detailed calculation see the Appendix), that is
\begin{align}\label{leading2}
\tfrac{1}{\eta} D_cD_d \tilde \chi&\hat=-\frac{4x_cx_d}{r^4\eta_\text{ret}}\partial^2_\eta\mathbb{T}(\eta_\text{ret})-\frac{4x^ax^bx_cx_d}{r^6} \partial^2_\eta\mathbb{T}_{ab}(\eta_\text{ret})+\mathcal{O}(H^3),
\end{align}
and
\begin{align}\label{leading3}
\frac{1}{\eta}\Big(\partial_\eta^2-\tfrac{1}{\eta}\partial_\eta \Big)\chi_{cd}\hat = -\frac{4}{r}\partial_r\bigg( \frac{1}{r}\partial_r\mathbb{T}_{cd}(\eta_\text{ret})\bigg).
\end{align}
Consequently, by combining expressions (\ref{leadingMiddleElectricWeyl}), (\ref{leading2}) and (\ref{leading3}), and using the identity $\partial_\eta \mathbb{T}_{ab}(\eta_\text{ret})=-\partial_r \mathbb{T}_{ab}(\eta_\text{ret})$ we calcualate the perturbed electric part of the Weyl tensor and write it explicitly up linear terms in $H$
\begin{align}
\mathcal{E}_{ab}
&= \bigg( \mathring q_a^c \mathring q_b^d - \tfrac{1}{3}\mathring q_{ab} \mathring q^{cd}\bigg) \bigg[-\frac{x^mx^nx_c x_d}{Hr^6}\partial_r^2 \mathbb{T}_{mn}(\eta_\text{ret}) -\frac{x_cx_d}{Hr^4}\partial_r^2\mathbb{T}(\eta_\text{ret})  d\eta'\nonumber\\
    &+\frac{2x^mx_{c}}{Hr^4}\partial^2_r\mathbb{T}_{dm}(\eta_\text{ret})+\frac{2x^mx_{d}}{Hr^4}\partial^2_r\mathbb{T}_{cm}(\eta_\text{ret})-\frac{2}{Hr^2} \partial^2_r\mathbb{T}_{cd}(\eta_\text{ret})\bigg]+\mathcal{O}(H^2).
\end{align}
It is also convenient to compute the following expression
\begin{align}    \mathcal{L}_T\chi_{ab}+2H\chi_{ab}&=-H(\eta\partial_\eta+r\partial_r)\chi_{ab}\nonumber\\
& \ \hat=\frac{4H}{r}\mathbb{T}_{ab}(\eta_\text{ret}).
\end{align}
The formula of (ABK) for energy emitted by gravitational waves now simplifies to
\begin{align}\label{semiAshtekarQF}
E_T &\hat= \frac{1}{4\pi} \int_{\mathcal{I}^+} dr d\Omega \ \bigg( \mathring q^{ec} \mathring q^{df} - \tfrac{1}{3}\mathring q^{ef} \mathring q^{cd}\bigg) \bigg[-\frac{\tilde x^m\tilde x^n\tilde x_c \tilde x_d}{Hr}\partial_r^2 \mathbb{T}_{mn}(\eta_\text{ret}) -\frac{\tilde x_c \tilde x_d}{Hr}\partial_r^2\mathbb{T}(\eta_\text{ret})   \nonumber\\
    &+\frac{4\tilde x^m\tilde x_{c}}{Hr}\partial^2_r\mathbb{T}_{dm}(\eta_\text{ret})-\frac{2}{Hr} \partial^2_r\mathbb{T}_{cd}(\eta_\text{ret})\bigg]\mathbb{T}_{ef}(\eta_\text{ret}) +\mathcal{O}(H^2)\nonumber\\
    &\hat= \frac{1}{4\pi} \int_{\mathcal{I}^+}  \frac{dr d\Omega}{Hr} \ \bigg( \mathring q^{ec} \mathring q^{df} - \tfrac{1}{3}\mathring q^{ef} \mathring q^{cd}\bigg) \bigg[\tilde x^m\hat x^n\tilde x_c \tilde x_d\partial_r \mathbb{T}_{mn}(\eta_\text{ret}) +\tilde x_c \hat x_d\partial_r\mathbb{T}(\eta_\text{ret}) \nonumber\\
    &-4\tilde x^m\tilde x_{c}\partial_r\mathbb{T}_{dm}(\eta_\text{ret})+2 \partial_r\mathbb{T}_{cd}(\eta_\text{ret})\bigg]\partial_r\mathbb{T}_{ef}(\eta_\text{ret}) +\mathcal{O}(H^2),
\end{align}
where we again used the notion (\ref{xTilde}), integrated by parts and kept only terms at most linear in $H$. Recall, that $T$ is the affine parameter along the integral curves of $T^a$ and satisfies $dT=-dr/(rH)$ at $\mathcal{I}^+$, in particular we may use $T$ as a new coordinate defined by the transformation
\begin{align}
    r\hat= \frac{1}{H}e^{-HT}.
\end{align}
Finally, integrating over the angular coordinates of (\ref{semiAshtekarQF}) yields
\begin{align}
E_T&=\frac{2}{15}\int dT  \ \bigg(4(\partial_r \mathbb{T}_{xx})^2+4(\partial_r \mathbb{T}_{yy})^2+4(\partial_r \mathbb{T}_{zz})^2+12(\partial_r \mathbb{T}_{xy})^2+12(\partial_r \mathbb{T}_{xz})^2+12(\partial_r \mathbb{T}_{yz})^2\\
    &-4\partial_r \mathbb{T}_{xx}\partial_r \mathbb{T}_{yy}(\eta_\text{ret})-4\partial_r \mathbb{T}_{xx}\partial_r \mathbb{T}_{zz}(\eta_\text{ret})-4\partial_r \mathbb{T}_{yy}\partial_r \mathbb{T}_{zz} \bigg)+\mathcal{O}(H^2).
\end{align}
where
\begin{align}
\partial_r\mathbb{T}_{ab}&
  \hat =-\frac{1}{2}\bigg[ \partial_T^3 Q_{ab}^{(\rho)}-3H \partial_T^2 Q_{ab}^{(\rho)}+H\partial_T^2 Q_{ab}^{(p)}\bigg]+\mathcal{O}(H^2).
\end{align}
Thus, we obtain the generalized quadrupole formula at $\mathcal{I}^+$ written up to the terms linear in the Hubble parameter $H$ (or equivalently $\sqrt{\Lambda/3}$)
\begin{align}\label{energyfinal0}
E_T&=\tfrac{2}{15}\int dT   \Bigg[(\partial_T^3 Q_{xx}^{(\rho)})^2+(\partial_T^3 Q_{yy}^{(\rho)})^2+(\partial_T^3 Q_{zz}^{(\rho)})^2+3(\partial_T^3 Q_{xy}^{(\rho)})^2+3(\partial_T^3 Q_{xz}^{(\rho)})^2+3(\partial_T^3 Q_{yz}^{(\rho)})^2\nonumber\\
&-(\partial_T^3 Q_{xx}^{(\rho)})(\partial_T^3 Q_{yy}^{(\rho)})-(\partial_T^3 Q_{xx}^{(\rho)})(\partial_T^3 Q_{zz}^{(\rho)})-(\partial_T^3 Q_{yy}^{(\rho)})(\partial_T^3 Q_{zz}^{(\rho)})\nonumber\\
&+2H\partial_T^3 Q_{xx}^{(\rho)}\Big(\partial_T^2 Q_{xx}^{(p)}-3\partial_T^2 Q_{xx}^{(\rho)}\Big)+2H\partial_T^3 Q_{yy}^{(\rho)}\Big(\partial_T^2Q_{yy}^{(p)}-3\partial_T^2 Q_{yy}^{(\rho)}\Big)\nonumber\\
&+2H\partial_T^3 Q_{zz}^{(\rho)}\Big(\partial_T^2Q_{zz}^{(p)}-3\partial_T^2 Q_{zz}^{(\rho)}\Big)+6H\partial_T^3 Q_{xy}^{(\rho)}\Big(\partial_T^2Q_{xy}^{(p)}-3\partial_T^2 Q_{xy}^{(\rho)}\Big)\nonumber\\
&+6H\partial_T^3 Q_{xz}^{(\rho)}\Big(\partial_T^2Q_{xz}^{(p)}-3\partial_T^2 Q_{xz}^{(\rho)}\Big)+6H\partial_T^3 Q_{yz}^{(\rho)}\Big(\partial_T^2Q_{yz}^{(p)}-3\partial_T^2 Q_{yz}^{(\rho)}\Big)\nonumber\\
&-H\partial_T^3 Q_{xx}^{(\rho)}\Big(\partial_T^2Q_{yy}^{(p)}-3\partial_T^2 Q_{yy}^{(\rho)}\Big)-H\partial_T^3 Q_{yy}^{(\rho)}\Big(\partial_T^2Q_{xx}^{(p)}-3\partial_T^2 Q_{xx}^{(\rho)}\Big)\nonumber\\
&-H\partial_T^3 Q_{xx}^{(\rho)}\Big(\partial_T^2Q_{zz}^{(p)}-3\partial_T^2 Q_{zz}^{(\rho)}\Big)-H\partial_T^3 Q_{zz}^{(\rho)}\Big(\partial_T^2Q_{xx}^{(p)}-3\partial_T^2 Q_{xx}^{(\rho)}\Big)\nonumber\\
&-H\partial_T^3 Q_{yy}^{(\rho)}\Big(\partial_T^2Q_{zz}^{(p)}-3\partial_T^2 Q_{zz}^{(\rho)}\Big)-H\partial_T^3 Q_{zz}^{(\rho)}\Big(\partial_T^2Q_{yy}^{(p)}-3\partial_T^2 Q_{yy}^{(\rho)}\Big)\Bigg]+\mathcal{O}(H^2).
\end{align}
which again may be written in terms of traceless quadrupole moments via (\ref{tracelessQM}) and the final expression is of the form:
\begin{align}\label{ABKquadfinal}
E_T=\tfrac{1}{45}\int dT\sum_{i,j=1}^3\Bigg[\bigg( \frac{d^3q^{(\rho)}_{ij}}{dT^3}\bigg)^2+2H\bigg( \frac{d^3q^{(\rho)}_{ij}}{dT^3}\frac{d^2q^{(p)}_{ij}}{dT^2}\bigg)\Bigg]+\mathcal{O}(H^2).
\end{align}
Therefore, the above energy flux of gravitational radiation across future infinity $\mathcal{I}^+$ manifestly agrees with the other two formulas, (\ref{quadrupoleDDR&JLsimlified}) and (\ref{quadrupoleHV}), computed on the cosmological horizon $\mathcal{H}$ via (DRL) and (HV) approaches.

\section{Summary}
We have investigated three different derivations of the generalizations of the quadrupole formula. In (DRL) approach \cite{DDR&JL}, it was calculated on cosmological horizon $\mathcal{H}$ instead of the conformal boundary used in radiation theory on the background of Minkowski spacetime and written explicitly up to terms linear in $\sqrt{\Lambda}$. Its form allows to take a limit as $\Lambda$ goes to zero simply by setting the Hubble parameter $H=0$ and the result recovers the famous Einstein's quadrupole formula. Our goal was two express the other to generalizations of the quadrupole formula by (HV) \cite{HV} and (ABK) \cite{ABK3} via the expansion in $\sqrt{\Lambda}$ and calculate the zeroth and first order terms. Even though those two quadrupole formulas, (\ref{quadrupoleHV}) and (\ref{quadrupoleTTABK}), look very similar, they come from conceptually different derivations and it requires some tedious calculations to show that they do coincide, at least up to terms linear in $\sqrt{\Lambda}$. Expression (\ref{quadrupoleHV}) is written in a tt-projection and evaluated over the cosmological horizon, whereas (\ref{quadrupoleTTABK}) was calculated in a TT-gauge at future infinity $\mathcal{I^+}$. The two notions, tt-projection and TT-gauge, are generically distinct, therefore, should not be used interchangeably. Even though, they do coincide at null infinity $\mathcal{I}^+$ of asymptotically flat spacetimes and as recently showed in \cite{HA} they also yield the same result for the energy flux computations for binary system in de Sitter background \cite{HA}, it is the TT-gauge that is the correct approach to extract transverse-traceless part of rank-2 symmetric tensor. Nevertheless, we expand the two expression for the radiated energy of the compact source and put them in an analogous form as (\ref{quadrupoleDDR&JLsimlified}) found in \cite{DDR&JL}. Therefore, each of them written up to linear terms in $\sqrt{\Lambda}$, (\ref{HVquadFinal}) and (\ref{ABKquadfinal}), not only admits the correct limit for the vanishing cosmological constant $\Lambda \rightarrow 0$, but also agree with the (\ref{quadrupoleDDR&JLsimlified}), as expected due to the symplectic theory on the space of solutions to Einstein's equations. This shows, that all three approaches, (ABK), (HV) and (DRL), provide the same generalization of the quadrupole formula up to terms linear in $\sqrt{\Lambda}$ (or equivalently in Hubble parameter $H$). 

Notice, however, that the terms linear in $\sqrt{\Lambda}$ are proportional to the second derivatives of the traceless pressure quadrupole moment $q^{(p)}_{ab}$, and in studying physical systems of interest (such as a binary system) it is often set to zero. Therefore, the first correction to the quadrupole formula is negligible and higher order terms are necessary to study the impact of the expansion of the Universe on the radiated energy of the time changing compact system. Higher order terms were taken into consideration in \cite{BH}, where the radiated power of the inspiring binary on circular orbit has been calculated in the high-frequency limit (where the expansion of the Universe is neglected during the orbital cycles of the system). Even though the result showed that the standard Einstein quadrupole formula is sufficient for studying  binary systems in high-frequency approximation, one could go beyond this approximation and probe the cosmological constant $\Lambda$ by measuring radiated energy by in the form of gravitational waves due to the binaries of supermassive black holes. The high-frequency approximation does not hold for such systems, and the corrections due to the background curvature could then become non-negligible. Recent results on the search of low-frequency gravitational waves through radio pulsar timing \cite{nanograv} give hope for new insight for the study of supermassive black hole binaries and significance of cosmological constant in gravitational radiation theory for expanding Universe.

\vspace{8mm}

{\noindent{\bf Acknowledgements:}}
JL was supported by 2021/43/B/ST2/02950 of the Polish National Science Centre.

\section{APPENDIX}
Here, we present a more detailed calculation of the expansion of the generalization of the quadrupole formula derived in \cite{ABK3}. Finding the explicit expression for the electric part of the Weyl tensor $\mathcal{E}_{ab}$ in terms of the metric perturbation $\chi_{ab}$ requires calculating $\tilde \chi$ from
\begin{align}
\tilde \chi &= \eta \int^\eta_{-\infty} \frac{\mathring D^a\chi_a+\partial_{\eta'} \chi}{\eta'} d\eta'.
\end{align}
We start with
\begin{align}
\chi_b &= -\frac{4x^a\eta^2}{r^3} \int^\eta_{-\infty} \frac{1}{\eta'^2_\text{ret}}\mathbb{T}_{ab}(\eta'_\text{ret})d\eta' -\frac{4x^a\eta}{r^2{\eta_\text{ret}}} \mathbb{T}_{ab}(\eta_\text{ret})
\end{align}
and calcualte its derivative
\begin{align}\label{append1}
D^b\chi_b &= -\frac{4\delta^{ab}\eta^2}{r^3} \int^\eta_{-\infty} \frac{1}{\eta'^2_\text{ret}}\mathbb{T}_{ab}(\eta'_\text{ret})d\eta'+\frac{12x^ax^b\eta^2}{r^5} \int^\eta_{-\infty} \frac{1}{\eta'^2_\text{ret}}\mathbb{T}_{ab}(\eta'_\text{ret})d\eta'-\frac{4\delta^{ab}\eta}{r^2{\eta_\text{ret}}} \mathbb{T}_{ab}(\eta_\text{ret})\nonumber\\
&+\frac{12x^ax^b\eta}{r^4{\eta_\text{ret}}} \mathbb{T}_{ab}(\eta_\text{ret})+\frac{4x^ax^b\eta}{r^3{\eta_\text{ret}}} \partial_\eta\mathbb{T}_{ab}(\eta_\text{ret}).
\end{align}
Next, using the expression (\ref{solution}) we compute:
\begin{align}\label{append2}
\partial_\eta\chi &= \frac{4}{r}\partial_\eta\mathbb{T}(\eta_\text{ret})+  4\frac{1}{\eta_\text{ret}} \partial_{\eta}\mathbb{T}(\eta_\text{ret})\nonumber\\
&= \frac{4\eta}{r\eta_\text{ret}}\partial_\eta\mathbb{T}(\eta_\text{ret}).
\end{align}
Then, combining eqs. (\ref{append1}) and (\ref{append2}) yields:
\begin{align}
 \frac{\mathring D^a\chi_a+\partial_{\eta'} \chi}{\eta'} &= -\frac{4\delta^{ab}\eta}{r^3} \int^\eta_{-\infty} \frac{1}{\eta'^2_\text{ret}}\mathbb{T}_{ab}(\eta'_\text{ret})d\eta'+\frac{12x^ax^b\eta}{r^5} \int^\eta_{-\infty} \frac{1}{\eta'^2_\text{ret}}\mathbb{T}_{ab}(\eta'_\text{ret})d\eta'-\frac{4\delta^{ab}}{r^2{\eta_\text{ret}}} \mathbb{T}_{ab}(\eta_\text{ret})\nonumber\\
&+\frac{12x^ax^b}{r^4{\eta_\text{ret}}} \mathbb{T}_{ab}(\eta_\text{ret})+\frac{4x^ax^b}{r^3{\eta_\text{ret}}} \partial_\eta\mathbb{T}_{ab}(\eta_\text{ret})+\frac{4}{r\eta_\text{ret}}\partial_\eta\mathbb{T}(\eta_\text{ret}).
\end{align}
Taking the derivative $\mathring D_d$ of the above gives
\begin{align}
\mathring D_d \frac{\mathring D^a\chi_a+\partial_{\eta'} \chi}{\eta'} &= \frac{12x_d\eta}{r^5} \int^\eta_{-\infty} \frac{1}{\eta'^2_\text{ret}}\mathbb{T}(\eta'_\text{ret})d\eta'+\frac{12x_d}{r^4{\eta_\text{ret}}} \mathbb{T}(\eta_\text{ret})+\frac{4x_d}{r^2\eta^2_\text{ret}}\partial_\eta\mathbb{T}(\eta_\text{ret})-\frac{4x_d}{r^2\eta_\text{ret}}\partial^2_\eta\mathbb{T}(\eta_\text{ret})\nonumber\\
&+\frac{24\delta^a_dx^b\eta}{r^5} \int^\eta_{-\infty} \frac{1}{\eta'^2_\text{ret}}\mathbb{T}_{ab}(\eta'_\text{ret})d\eta'-\frac{60x^ax^bx_d\eta}{r^7} \int^\eta_{-\infty} \frac{1}{\eta'^2_\text{ret}}\mathbb{T}_{ab}(\eta'_\text{ret})d\eta'-\frac{60x^ax^bx_d}{r^6{\eta_\text{ret}}}\mathbb{T}_{ab}(\eta_\text{ret})\nonumber\\
&+\frac{24\delta^a_dx^b}{r^4{\eta_\text{ret}}} \mathbb{T}_{ab}(\eta_\text{ret})+\frac{8\delta_d^ax^b}{r^3{\eta_\text{ret}}} \partial_\eta\mathbb{T}_{ab}(\eta_\text{ret})+\frac{4(-6\eta+7r)x^ax^bx_d}{r^5{\eta^2_\text{ret}}} \partial_\eta\mathbb{T}_{ab}(\eta_\text{ret})\nonumber\\
&-\frac{4x^ax^bx_d}{r^4{\eta_\text{ret}}} \partial^2_\eta\mathbb{T}_{ab}(\eta_\text{ret}).
\end{align}
Then, by computing yet another derivative of the above we find
\begin{align}
D_cD_d \frac{\mathring D^a\chi_a+\partial_{\eta'} \chi}{\eta'} &=\frac{12\delta_{cd}\eta}{r^5} \int^\eta_{-\infty} \frac{1}{\eta'^2_\text{ret}}\mathbb{T}(\eta'_\text{ret})d\eta'-\frac{60x_cx_d\eta}{r^7} \int^\eta_{-\infty} \frac{1}{\eta'^2_\text{ret}}\mathbb{T}(\eta'_\text{ret})d\eta'-\frac{60x_cx_d}{r^6{\eta_\text{ret}}}\mathbb{T}(\eta_\text{ret})\nonumber\\
&+\frac{12\delta_{cd}}{r^4{\eta_\text{ret}}} \mathbb{T}(\eta_\text{ret})+\frac{4x_cx_d\eta(-3\eta+4r)}{r^5{\eta^3_\text{ret}}}\partial_\eta \mathbb{T}(\eta_\text{ret})+\frac{4\delta_{cd}}{r^2\eta^2_\text{ret}}\partial_\eta\mathbb{T}(\eta_\text{ret})\nonumber\\
&+\frac{4x_cx_d(2\eta-4r)}{r^4\eta^2_\text{ret}}\partial^2_\eta\mathbb{T}(\eta_\text{ret})-\frac{4\delta_{cd}}{r^2\eta_\text{ret}}\partial^2_\eta\mathbb{T}(\eta_\text{ret})+\frac{4x_cx_d}{r^3\eta_\text{ret}}\partial^3_\eta\mathbb{T}(\eta_\text{ret})\nonumber\\
&+\frac{24\eta}{r^5} \int^\eta_{-\infty} \frac{1}{\eta'^2_\text{ret}}\mathbb{T}_{cd}(\eta'_\text{ret})d\eta'-\frac{120x_cx^b\eta}{r^7} \int^\eta_{-\infty} \frac{1}{\eta'^2_\text{ret}}\mathbb{T}_{bd}(\eta'_\text{ret})d\eta'\nonumber\\
&-\frac{120x^bx_d\eta}{r^7} \int^\eta_{-\infty} \frac{1}{\eta'^2_\text{ret}}\mathbb{T}_{cb}(\eta'_\text{ret})d\eta'-\frac{60x^ax^b\delta_{cd}\eta}{r^7} \int^\eta_{-\infty} \frac{1}{\eta'^2_\text{ret}}\mathbb{T}_{ab}(\eta'_\text{ret})d\eta'\nonumber\\
&+\frac{420x^ax^bx_cx_d\eta}{r^9} \int^\eta_{-\infty} \frac{1}{\eta'^2_\text{ret}}\mathbb{T}_{ab}(\eta'_\text{ret})d\eta'+\frac{420x^ax^bx_cx_d}{r^8{\eta_\text{ret}}} \mathbb{T}_{ab}(\eta_\text{ret})\nonumber\\
&-\frac{120x^bx_d}{r^6{\eta_\text{ret}}}\mathbb{T}_{cb}(\eta_\text{ret})-\frac{120x^bx_c}{r^6{\eta_\text{ret}}} \mathbb{T}_{bd}(\eta_\text{ret})-\frac{60x^ax^b\delta_{cd}}{r^6{\eta_\text{ret}}}\mathbb{T}_{ab}(\eta_\text{ret})\nonumber\\
&+\frac{4(45\eta^2-100\eta r+57r^2)x^ax^bx_cx_d}{r^7{\eta^3_\text{ret}}}\partial_\eta\mathbb{T}_{ab}(\eta_\text{ret})+\frac{24}{r^4{\eta_\text{ret}}} \mathbb{T}_{cd}(\eta_\text{ret})\nonumber\\
&+\frac{8}{r^3{\eta_\text{ret}}} \partial_\eta\mathbb{T}_{cd}(\eta_\text{ret})+\frac{8(-6\eta+7r)x^bx_c}{r^5{\eta^2_\text{ret}}} \partial_\eta\mathbb{T}_{bd}(\eta_\text{ret})\nonumber\\
&+\frac{8(-6\eta+7r)x^bx_d}{r^5{\eta^2_\text{ret}}} \partial_\eta\mathbb{T}_{cb}(\eta_\text{ret})+\frac{4(-6\eta+7r)x^ax^b\delta_{cd}}{r^5{\eta^2_\text{ret}}} \partial_\eta\mathbb{T}_{ab}(\eta_\text{ret})\nonumber\\
&+\frac{8(3\eta-4r)x^ax^bx_cx_d}{r^6{\eta^2_\text{ret}}} \partial^2_\eta\mathbb{T}_{ab}(\eta_\text{ret})-\frac{8x^bx_d}{r^4{\eta_\text{ret}}} \partial^2_\eta\mathbb{T}_{cb}(\eta_\text{ret})-\frac{8x^bx_c}{r^4{\eta_\text{ret}}} \partial^2_\eta\mathbb{T}_{bd}(\eta_\text{ret})\nonumber\\
&-\frac{4x^ax^b\delta_{cd}}{r^4{\eta_\text{ret}}} \partial^2_\eta\mathbb{T}_{ab}(\eta_\text{ret})+\frac{4x^ax^bx_cx_d}{r^5{\eta_\text{ret}}} \partial^3_\eta\mathbb{T}_{ab}(\eta_\text{ret}).
\end{align}
Finally, we integrate the above expression, divide it by $\eta$ and find its value at conformal infinity of de Sitter spacetime $\mathcal{I}^+$ by setting $\eta=0$:
\begin{align}
  D_cD_d \tilde \chi&=\eta \int^\eta_{-\infty}D_cD_d \frac{\mathring D^a\chi_a+\partial_{\eta'} \chi}{\eta'}d\eta'\nonumber\\
  &\hat=-\frac{4x_cx_d}{r^4\eta_\text{ret}}\partial^2_\eta\mathbb{T}(\eta_\text{ret})-\frac{4x^ax^bx_cx_d}{r^6} \partial^2_\eta\mathbb{T}_{ab}(\eta_\text{ret})+\mathcal{O}(H^3).
\end{align}

 \thispagestyle{empty}

\end{document}